\documentclass[prd,reprint,twocolumn,nofootinbib,showpacs]{revtex4-1}
\usepackage{bm,amssymb,amsmath,graphicx}
\usepackage{bm,hyperref,amssymb}

\newcommand{\bA}{{\mathbf A}}
\newcommand{\bB}{{\mathbf B}}

\begin{document}
\title{Tails of plane wave spacetimes: Wave-wave scattering in general relativity}
\author{Abraham I. Harte}
\affiliation{Max-Planck-Institut f\"{u}r Gravitationsphysik, Albert-Einstein-Institut,
\\
Am M\"{u}hlenberg 1, 14476 Golm, Germany}
\email{Electronic address: harte@aei.mpg.de}

\begin{abstract}
One of the most important characteristics of light in flat spacetime is that it satisfies Huygens' principle: Initial data for the vacuum Maxwell equations evolves sharply along null (and not timelike) geodesics. In flat spacetime, there are no tails which linger behind expanding wavefronts. Tails generically do exist, however, if the background spacetime is curved. The only non-flat vacuum geometries where electromagnetic fields satisfy Huygens' principle are known to be those associated with gravitational plane waves. This paper investigates whether perturbations to the plane wave geometry itself also propagate without tails. First-order perturbations to all locally-constructed curvature scalars are indeed found to satisfy Huygens' principles. Despite this, gravitational tails do exist. Locally, they can only perturb one plane wave spacetime into another plane wave spacetime. A weak localized beam of gravitational radiation passing through an arbitrarily-strong plane wave therefore leaves behind only a slight perturbation to the waveform of the background plane wave. The planar symmetry of that wave cannot be disturbed by any linear tail. These results are obtained by first deriving the retarded Green function for Lorenz-gauge metric perturbations and then analyzing its consequences for generic initial-value problems.
\end{abstract}


\maketitle

\section{Introduction}

In flat spacetime and in the absence of intervening matter, electromagnetic fields do not develop tails. Light propagates without dispersion and bursts of radiation which are initially sharp remain sharp as they evolve. More precisely, the electromagnetic field $F_{ab}(x)$ at an event $x$ depends on current densities and initial data only on, and not inside, the past null cone of $x$. This is referred to as Huygens' principle. 

It is fundamental to our everyday lives that light and sound as we tend to experience them satisfy at least approximate versions of Huygens' principle. If this were not true, moving objects would appear to blur together and spoken language would be impractical. There are, however, many wave phenomena for which Huygens' principle does fail. A pebble falling into a pond, for example, produces water waves which persist at any fixed location long after the initial wavefront has passed. Whether or not Huygens' principle is valid for any particular system may be interpreted as a property of the initial-value problem associated with the particular differential equations used to describe that system. From this point of view, only a handful of reasonable wave equations satisfy Huygens' principle exactly. The vast majority of possibilities develop tails.

In the context of general relativity, it is of particular interest to understand how the curvature of spacetime is related to the presence of tails. The vacuum Maxwell equations provide a particularly well-studied example. Restricting to four-dimensional background spacetimes satisfying the vacuum Einstein equation $R_{ab} = 0$, Huygens' principle holds if and only if the geometry is either flat or is associated with a plane-symmetric gravitational wave \cite{GuentherBook, OnlyPlaneWavesEM}. Slightly more can be said by noting that Maxwell's equations are conformally invariant. Electromagnetic waves therefore propagate tail-free on all (not necessarily vacuum) metrics which are conformal either to flat spacetime or to a plane wave. If any further possibilities exist, they are highly constrained by the results reviewed in \cite{HuygensReview1, HuygensReview2, GuentherBook}.

It is clear that plane wave spacetimes play a privileged role in the theory of electromagnetic fields. This is not, it should be emphasized, a peculiarity of Maxwell's equations. The massless Klein-Gordon equation also satisfies Huygens' principle in flat and plane wave spacetimes \cite{Guenther}, but essentially no others \cite{OnlyPlaneWavesScalar}. Massless spin-$\frac{1}{2}$ fields propagate tail-free on plane wave backgrounds as well \cite{WeylEqn, HuygensReview1}.

This paper explores whether perturbations to the plane wave spacetimes themselves satisfy a version of Huygens' principle. In physical terms, does a localized packet of gravitational waves passing through a background plane wave leave behind a ``wake?'' The answer is that such a wake does exist, although its form is remarkably simple.

There are many motivations of studying plane wave spacetimes. Their most obvious application is as exact models of gravitational radiation in general relativity. Separate from this, plane wave spacetimes also have a number of unusual mathematical properties. Their relevance for (non-gravitational) wave tails has already been mentioned. Additionally, Penrose has shown \cite{Penrose} that plane waves arise as universal limits of all spacetimes in the vicinity of null geodesics. This might imply that certain aspects of ``ultra-relativistic physics'' in generic curved spacetimes can be understood using effective plane wave backgrounds. Ideas of this type have already been applied to understand properties of classical \cite{HarteCaustics} and quantum \cite{Hollowood} fields propagating on curved backgrounds. The Penrose limit and other features of plane wave spacetimes have also led to applications in string theory and related topics \cite{String1, String2}.

Regardless of motivation, we consider linear metric perturbations on vacuum plane wave backgrounds in four spacetime dimensions. The few existing results on Huygens' principles in linearized general relativity cannot be used to immediately answer whether gravitational tails exist on plane wave backgrounds. Waylen \cite{Waylen} considered the linearized Einstein equation on general backgrounds and in Lorenz gauge. He then asked whether the trace $g^{ab}(x) h_{ab}(x)$ of the first-order metric perturbation $h_{ab}$ at a point $x$ depended on initial data lying inside the past null cone of $x$. This was always found to occur unless the background spacetime is either flat or a plane wave. Later, Noonan \cite{GRHuygensNoonan} and W\"{u}nsch \cite{GRHuygensWuensch} separated asked -- again in Lorenz gauge -- whether \textit{any} part of $h_{ab}(x)$ depended on initial data inside the past null cone of $x$. They found that this always occurred unless the background was flat.

None of these results settle the issue.  Even if a metric perturbation develops a tail in Lorenz gauge, it might not do so in another gauge. This is precisely what occurs for electromagnetic fields on plane wave backgrounds. In that case, the vector potential in Lorenz gauge develops a tail (as it does in all curved spacetimes \cite{NoonanVect}). Such tails can, however, be removed by appropriate gauge transformations \cite{Kuenzle}. The electromagnetic field $F_{ab} = 2 \nabla_{[a} A_{b]}$ satisfies Huygens' principle even though $A_a$ typically does not. 

The problem of gauge is significantly more subtle for metric perturbations. This is both because the class of possible gauge transformations is larger than in electromagnetism and also because it is less clear which observables should be considered. The metric at a single point is not very interesting, for example. One might instead consider perturbations to curvature scalars such as $R^{abcd} R_{abcd}$. These scalars vanish on the background, so their first-order perturbations are gauge-invariant. It is shown here that all such perturbations satisfy Huygens' principles: Every locally-constructed polynomial in the curvature and its derivatives propagates tail-free on plane wave backgrounds. Despite this, there is a sense in which gravitational tails -- unlike their scalar or electromagnetic counterparts -- do exist. Locally, these can only perturb one plane wave spacetime into another plane wave spacetime. Even initial data with no symmetries whatsoever produces tails which retain the full\footnote{Some plane wave spacetimes admit a sixth Killing field. This is not necessarily preserved by tails.} five-dimensional space of Killing fields associated with the background. The waveform associated with that background may be perturbed, however. This is found to depend on only a single component of the initial metric perturbation. 

Our results are derived by first considering the general problem of first-order metric perturbations on plane wave backgrounds. We derive the retarded Green function for metric perturbations in Lorenz gauge and then use a Kirchhoff integral to write an arbitrary metric perturbation in terms of Cauchy data and any sources which may be present. The tail of the resulting Green function is used to derive various properties satisfied by the ``tail portion of the metric'' in Lorenz gauge. Lastly, an explicit gauge transformation is constructed which transforms the perturbed metric into the canonical form for a plane wave spacetime. 

Sec. \ref{Sect:PlaneWaves} briefly reviews some properties of plane wave spacetimes needed for the later development. Sec. \ref{Sect:EM} then illustrates how simple test fields propagate on plane wave backgrounds by deriving Green functions for scalar and electromagnetic fields. While the main conclusions of this section have appeared elsewhere in various forms, they are re-derived here more succinctly and to allow for a more direct comparison to the gravitational case. The methods reviewed in Sec. \ref{Sect:EM} are applied in new ways in Sec. \ref{Sect:GravGreen} to derive a Green function for metric perturbations on plane wave backgrounds. This Green function has a tail which is interpreted in Sec. \ref{Sect:PlaneToPlane}.

$3+1$ dimensions are assumed throughout this paper. The signature is taken to be $+2$ and the sign of the Riemann tensor is defined such that $R_{abc}{}^{d} \omega_d = 2 \nabla_{[a} \nabla_{b]} \omega_c$ for any $\omega_a$. Latin letters $a,b,\ldots$ from the beginning of the alphabet are used to denote abstract indices. The letters $i,j, \ldots$ instead refer to coordinates $x^i$ ($i=1,2$) which are transverse to the direction of propagation associated with the background plane wave. The coordinates $x^i$ are not to be confused with spacetime points denoted by $x,y,\ldots$ (without indices).

%
%
%

%
%

\section{Plane wave spacetimes}
\label{Sect:PlaneWaves}

Various aspects of plane wave spacetimes have been reviewed in \cite{EhlersKundt, GlobalLorentz, HarteCaustics, HarteLensing}. For our purposes, these are most naturally introduced as special cases of the plane-fronted waves with parallel rays (pp-waves). Although the concept is more general, we focus only on pp-waves which are solutions to the vacuum Einstein equation $R_{ab} =0$ in 4 spacetime dimensions. It is then sufficient to consider a simply connected region $\mathcal{M}$ which admits a vector field $\ell^a$ that is everywhere null, non-vanishing, and covariantly constant \cite{EhlersKundt}:
\begin{equation}
    \ell^a \ell_a = \nabla_b \ell^a = 0.
    \label{ellConds}
\end{equation}
The integral curves of $\ell^a$ are interpreted as the rays of the gravitational wave. These are free of expansion, shear, and twist. There is a sense in which all such rays remain parallel to one another and are orthogonal to a family of planar wavefronts. It follows from the simple-connectedness of $\mathcal{M}$ and from $\nabla_{[a} \ell_{b]} = 0$ that a scalar field $u$ exists such that
\begin{equation}
    \ell_a = - \nabla_a u. 
    \label{uDef}
\end{equation}
$u$ describes the phase of the gravitational wave. Hypersurfaces of constant $u$ are null and are referred to as wave surfaces. 

Any vacuum spacetime which admits a constant null vector field $\ell^a$ also admits a non-vanishing and covariantly constant null 2-form $f_{ab} = f_{[ab]}$ \cite{EhlersKundt}. $f_{ab}$ and its dual $f^*_{ab} = \epsilon_{ab}{}^{cd} f_{cd}/2$ (which is also null and constant) describe the plane of polarization associated with the gravitational wave. This is always transverse to the direction of propagation in the sense that $f_{ab} \ell^b = f^*_{ab} \ell^b= 0$.

An explicit metric for pp-wave spacetimes may be constructed using a phase coordinate $u$ which satisfies \eqref{uDef}, an affine parameter $v$ defined by
\begin{equation}
    \ell^a = \frac{\partial}{\partial v},
    \label{ellDef}
\end{equation}
and two more scalar fields $x^i$ ($i=1,2$) related to $f_{ab}$ via
\begin{equation}
    f_{ab} = \ell_{[a} \nabla_{b]} x^1, \qquad f_{ab}^* = - \ell_{[a} \nabla_{b]} x^2.
    \label{fDef}
\end{equation}
These choices result in the so-called Brinkmann metric
\begin{equation}
    d s^2 = - 2 du dv + H(u,x^k) du^2 + (dx^1)^2 + (dx^2)^2
    \label{ppMetric}  
\end{equation}
for general pp-wave spacetimes. The wave profile $H$ which appears here is constrained by the vacuum Einstein equation to satisfy
\begin{equation}
    \nabla^2 H(u, x^i ) = 0,
    \label{ppVac}
\end{equation}
where $\nabla^2 = \partial_{x^1}^2+ \partial_{x^2}^2$ is usual Laplace operator in two dimensions. $H$ is otherwise arbitrary. 

The Brinkmann coordinates $(u,v,x^i)$ have a number of useful properties. Unlike coordinate systems which are more closely related to typical perturbative discussions of gravitational radiation in transverse-traceless gauge, Brinkmann coordinates do not develop coordinate singularities. Brinkmann coordinates are also harmonic and place the metric into the Kerr-Schild form
\begin{equation}
    g_{ab} = \eta_{ab} + H \ell_a \ell_b.
\end{equation}
Here, $\eta_{ab}$ is flat and $\ell_a = g_{ab} \ell^b = \eta_{ab} \ell^b$. The vector field $\ell^a$ is null with respect to both $\eta_{ab}$ and $g_{ab}$. Viewing $H \ell_a \ell_b$ as an arbitrarily-large perturbation on the ``background'' $\eta_{ab}$, \eqref{ppVac} implies that this perturbation satisfies the linearized \textit{as well as} the fully nonlinear Einstein equations (which is true for all Kerr-Schild metrics \cite{Xanth}). Indeed, pp-waves obey a type of exact linear superposition. This is typically not apparent in other coordinate systems.

In general, pp-waves admit only one Killing vector. Recall, however, that electromagnetic plane waves in flat spacetime are preserved by a five-dimensional space of Killing vector fields. Gravitational plane waves may therefore be defined as those pp-waves which admit at least five linearly independent Killing fields  \cite{Bondi}. This can be shown to imply that for a plane wave spacetime, $H(u,x^i)$ can be at most quadratic in the two transverse coordinates $x^i$. A coordinate transformation may always be used to eliminate all terms in $H(u,x^i)$ which are either independent of or linear in $x^i$. Plane wave spacetimes may therefore be described by a symmetric $2 \times 2$ matrix $H_{ij}(u)$ satisfying
\begin{equation}
    H(u , x^k) = H_{ij}(u) x^i x^j.
    \label{HtoH}
\end{equation}
Einstein's equation \eqref{ppVac} then reduces to the trivial algebraic constraint that $H_{ij}$ be trace-free. This leaves two free functions of $u$ corresponding to waveforms\footnote{This notion of waveform is different from the one typically used in perturbative descriptions of gravitational radiation. $H_{ij}$ directly describes the local curvature $R_{abc}{}^{d}$ [see \eqref{Weyl}]. The waveforms associated with perturbation theory in transverse-traceless gauge must instead be differentiated twice in order to recover $R_{abc}{}^{d}$.} associated with the two polarization states of the gravitational wave. 

It can be convenient to replace the real matrix $H_{ij}(u)$ by a single complex function $\mathcal{H}(u)$ via
\begin{equation}
    \mathcal{H} = H_{11} + i H_{12}. 
\end{equation}
Similarly complexifying the transverse coordinates using
\begin{equation}
    \xi = \frac{1}{\sqrt{2}} ( x^1 + i x^2),
\end{equation}
the pp-wave line element \eqref{ppMetric} with a vacuum plane wave-type profile \eqref{HtoH} is then equivalent to
\begin{equation}
    d s^2 = 2 \left( - d u dv + \Re \big[\mathcal{H}(u) \bar{\xi}^2 \big] d u^2 +  d \xi d \bar{\xi} \right).
    \label{PlaneWaveMetric}
\end{equation}
Here, overbars denote complex conjugates and $\Re$ returns the real component of its argument. Both the real and complex representations of plane wave spacetimes are used interchangeably below.

The complex waveform $\mathcal{H}$ constitutes a nearly gauge-invariant way to describe a vacuum plane wave spacetime. To see this, first note that vector fields $\ell^a$ satisfying \eqref{ellConds} are unique up to an overall constant. Phases $u$ satisfying \eqref{uDef} may therefore be altered only by constant affine transformations. $\xi$ and $\bar{\xi}$ can vary by rotations in the complex plane, but also by more subtle $u$-dependent ``translations'' which affect $v$ as well. All coordinate transformations that preserve the form of the Brinkmann metric \eqref{PlaneWaveMetric} are known \cite{EhlersKundt}. Applying them results in each plane wave spacetime being describable by a 3-parameter family of possible waveforms:
\begin{equation}
    \mathcal{H}( u ) \rightarrow a^2  \mathcal{H}( a u - b ) e^{i c}.
    \label{WaveformGauge}
\end{equation}
Here, $a \neq 0$, $b$, and $c$ are arbitrary real constants. These cannot change the overall ``shape'' of a non-constant waveform. Note, however, that the overall magnitude of a particular $\mathcal{H}$ is not invariant. There is no intrinsic notion for the amplitude of a gravitational plane wave unless additional structure is introduced (such as a preferred observer).

Without any such structure, it is difficult to find any local scalar invariants in plane wave spacetimes. All polynomials in the curvature and its derivatives vanish, for example. A nontrivial invariant may nevertheless be constructed from a particular $\mathcal{H}$ using \cite{EhlersKundt, PlaneWaveClassify}
\begin{equation}
    I(u) = \frac{ [\partial_u \ln \mathcal{H}(u) ]^2 }{ | \mathcal{H}(u) | }. 
\end{equation}
This is unique up to affine transformations of its argument. It is not, however, a reasonable measure of amplitude: $I=0$ when $\mathcal{H}$ is a nonzero constant even though this case is physically nontrivial.

Local physics in a curved spacetime is largely determined by the Riemann tensor $R_{abc}{}^{d}$. For plane waves, the only independent coordinate components of the curvature are determined by
\begin{equation}
    R_{uiuj} = - H_{ij}.
\end{equation}
It can be convenient to re-express this equation in terms of a complex null tetrad $(\ell^a, n^a, m^a, \bar{m}^a)$. Let $\ell^a$ be defined by \eqref{ellDef} and  \begin{align}
    n^a = \frac{\partial}{\partial u} + \Re (\mathcal{H} \bar{\xi}^2 ) \frac{\partial}{\partial v} , \quad     m^a = \frac{\partial}{\partial \bar{\xi}} .
    \label{TetradDef}
\end{align}
All scalar products of these vector fields vanish except for $\ell^a n_a = - 1$ and $m^a \bar{m}_a = 1$. The associated Weyl scalars vanish except for 
\begin{equation}
    \Psi_4 = C_{abcd} n^a \bar{m}^b n^c \bar{m}^d = - \bar{\mathcal{H}}.
    \label{Weyl}
\end{equation}
Derivatives of the tetrad are
\begin{equation}
    \nabla_a n^b = 2 \ell_a \Re( \bar{\mathcal{H}} \xi m^b) , \qquad \nabla_a m^b = \mathcal{H} \bar{\xi} \ell_a \ell^b,
    \label{TetradD}
\end{equation}
from which it follows that all divergences and d'Alembertians of the tetrad components must vanish (e.g., $\nabla_a n^a = 0$ and $\Box n^a = \nabla^b \nabla_b n^a = 0$).

\subsection*{Geodesics and associated bitensors}

Geodesics in plane wave spacetimes fall into one of two classes. The simpler (and less interesting) case comprises those geodesics which remain confined to a single  hypersurface of constant $u$. The only causal geodesics with this property are the null integral curves of $\ell^a$; the rays of the gravitational wave. More interesting are the geodesics which pass through wave surfaces. Recalling that $\ell^a$ is Killing, the $u$ coordinate may be used as an affine parameter for all such geodesics. All timelike and almost all null geodesics are in this class. Their transverse coordinates $\xi, \bar{\xi}$ satisfy
\begin{equation}
    \ddot{\xi} = \mathcal{H} \bar{\xi} ,
    \label{TransverseCoordsIm}
\end{equation}
where overdots denote derivatives with respect to $u$. Solutions to this equation directly determine almost all interesting properties of plane wave spacetimes. They describe not only the behaviors of individual geodesics, but also geodesic deviation, parallel transport, distances between finitely-separated points, and so on.

Relations of this type have previously been formulated using the real form 
\begin{equation}
    \ddot{x}_i = H_{ij} x^j
    \label{TransverseCoordsRe}
\end{equation}
of \eqref{TransverseCoordsIm} \cite{HarteCaustics}. This equation is linear, so its solutions depend linearly on initial conditions. Specifying these conditions at a phase $u'$, there exist $2 \times 2$ real matrices $A_{ij}$ and $B_{ij}$ such that
\begin{equation}
    x_i(u) = A_{ij} (u,u') x^j(u') + B_{ij}(u,u') \dot{x}^j(u').
    \label{xExpand}
\end{equation}
The two propagators here satisfy
\begin{subequations}
\begin{align}
    \partial_u^2 A_{ij}(u,u') = H_{ik}(u) A_{kj}(u,u')
    \\
    \partial_u^2 B_{ij}(u,u') = H_{ik}(u) B_{kj}(u,u')
\end{align}
\label{ABDef}
\end{subequations}
together with
\begin{align}
    [A_{ij}] = [\partial_u B_{ij}] = \delta_{ij}, \quad [\partial_u A_{ij}] = [B_{ij}] = 0.
    \label{ABInit}
\end{align}
Brackets are used in these last equations to denote the coincidence limit $\lim_{u \rightarrow u'}$. Knowledge of the waveform $H_{ij}$ (or equivalently $\mathcal{H}$) is sufficient to compute $A_{ij}$ and $B_{ij}$ everywhere. Detailed properties of these matrices are discussed in \cite{HarteCaustics, HarteLensing}.

Eq. \eqref{TransverseCoordsRe} is analogous to the equations which describe Newtonian masses coupled by springs with time-varying stiffnesses. That $H_{ij}$ is trace-free implies that some of these stiffnesses must be negative. In cases where $H_{ij}$ is periodic, \eqref{TransverseCoordsRe} is a form of Hill's equation. This arises in many areas of physics and engineering, and is known to have a rich phenomenology \cite{HillBook}. No particular form for $H_{ij}$ is assumed here, however (other than a sufficient degree of regularity).

It is intuitively clear that solutions to oscillator-like equations tend to eventually pass through zero. Fixing some $u'$, there generically exist some values of $U \neq u'$ such that $\det B_{ij}(U,u') = 0$. This means that in \eqref{xExpand}, the initial transverse velocity $\dot{x}^i(u')$ of a geodesic may be varied in the null space of $B_{ij}(U,u')$ without affecting the final transverse position $x^i(U)$. The initial and final $v$ coordinates of such a family of geodesics may also be arranged to coincide. Two points in a plane wave spacetime can therefore be connected by an infinite number of geodesics when $\det B_{ij} = 0$. This clearly signals the boundary of a normal neighborhood. It is also closely related to the largest domains on which initial data may be used to uniquely determine the future (since plane wave spacetimes are not globally hyperbolic \cite{PenrosePlane}). For simplicity, this paper restricts attention to regions which are sufficiently small that $\det B_{ij}(u,u') \neq 0$ for all relevant phases $u \neq u'$. Any pair of points is then connected by exactly one geodesic, there are no conjugate points, and initial-value problems associated with standard equations are well-posed.

Various geometric objects important for the construction of Green functions may be found explicitly in terms of $A_{ij}$ and $B_{ij}$ (see, e.g., \cite{HarteCaustics, Hollowood}). Synge's function $\sigma(x,x') = \sigma(x',x)$, which returns one-half of the squared geodesic distance between the points $x$ and $x'$, is
\begin{align}
    \sigma = \frac{1}{2} (u-u') [ - 2 (v-v') + (\partial_u \bB \bB^{-1})_{ij} x^i x^j
    \nonumber
    \\
    ~ + (\bB^{-1} \bA)_{ij} x'^i x'^j - 2 (\bB^{-1})_{ij} x'^i x^j] .
    \label{Sigma}
\end{align}
Here $(\bB^{-1})_{ij}$ denotes the matrix inverse of $B_{ij}$. By assumption, $(u-u') [\bB^{-1}(u,u')]_{ij}$ exists in all regions which are of interest here. $\sigma(x,x')$ is positive, negative, or zero if its arguments are spacelike-, timelike-, or null-separated.

Another important bitensor is the ``scalarized'' van Vleck determinant $\Delta(x,x') = \Delta(x',x)$. In plane wave spacetimes,
\begin{equation}
    \Delta (x,x') = \frac{ (u-u')^2 }{ \det B_{ij}(u,u') } .
    \label{VanVleck} 
\end{equation}
Holding $x'$ fixed, this is related to the expansion of the congruence of geodesics passing through $x'$ \cite{PoissonRev}. $\Delta$ depends only on the phase coordinates of its arguments in plane wave spacetimes.

The last important bitensor needed here is the parallel propagator $g^{a}{}_{a'}(x,x') = g_{a'}{}^{a} (x',x)$. Contracting this with any vector at $x'$ returns that same vector parallel-propagated to $x$ along the unique geodesic which passes through both $x$ and $x'$. Coordinate components of $g^{a}{}_{a'}$ have been computed in \cite{HarteCaustics}. Re-expressed in terms of the null tetrad \eqref{TetradDef},
\begin{align}
    g^{a}{}_{a'} = -\ell^a \big[ n_{a'} + \zeta \bar{\zeta} \ell_{a'} - 2 \Re(\bar{\zeta} m_{a'}) \big] - n^a \ell_{a'}
    \nonumber
    \\
    ~ + 2 \Re \big[ m^a ( \bar{m}_{a'} - \bar{\zeta} \ell_{a'}) \big] .
    \label{ParallelProp}
\end{align}
The complex function $\zeta(x,x')$ appearing here may be found by first considering a geodesic $y(\tau)$ which passes through two given points $x$ and $x'$. Using \eqref{xExpand} with the boundary conditions $y^i(u) = x^i$ and $y^i(u') = x'^i$, the transverse components of such a geodesic satisfy
\begin{align}
    y_i (\tau) =  [\bB(\tau,u') \bB^{-1}(u,u')]_{ij} [ x^j - A^{j}{}_{k}(u,u') x'^k ] 
    \nonumber
    \\
    ~ + A_{ij}(\tau,u') x'^j.
    \label{Geodesic}
\end{align}
Then $\zeta$ is given by
\begin{equation}
    \zeta = \frac{1}{\sqrt{2}} \left\{ \big[ \dot{y}^1(u) - \dot{y}^1(u') \big] + i \big[ \dot{y}^2(u) - \dot{y}^2(u') \big] \right\}.
    \label{ZetaDef}
\end{equation}
This is independent of $v,v'$ and linear in $x^i, x'^i$ (or equivalently $\xi, \bar{\xi}$). It also antisymmetric in its arguments: $\zeta(x,x') = - \zeta(x',x)$.

To summarize, the bitensors $\sigma$, $\Delta$, $g^{a}{}_{a'}$ are all trivial to compute once $A_{ij}$ and $B_{ij}$ are known. These latter matrices in turn depend only on the waveform $H_{ij}$. 


\section{Scalar and electromagnetic fields}
\label{Sect:EM}

Before considering metric perturbations propagating on a plane wave background, it is instructive to first review the behavior of scalar and electromagnetic fields in this context. As outlined above, we restrict attention only to regions which are sufficiently small that all initial value problems are well-posed and exactly one geodesic passes through any two distinct points. 

Consider a massless scalar field $\psi$ propagating on a plane wave background. Supposing the existence of a charge density $\rho$, this satisfies
\begin{equation}
    \Box \psi = \nabla^a \nabla_a \psi = - 4 \pi \rho.
    \label{ScalarField}
\end{equation}
It is convenient to introduce a retarded Green function $G(x,x')$. Let
\begin{equation}
    \Box G(x,x') = - 4\pi \delta(x,x'),
    \label{GScalar}
\end{equation}
and demand that $G(x,x')$ vanish when $x$ is not in the causal future of $x'$. Given an appropriate spacelike hypersurface $\Sigma$ on which to impose initial data, arbitrary solutions to \eqref{ScalarField} in the future of $\Sigma$ may be be written in terms of the Kirchhoff integral
\begin{align}
    & \psi(x) = \! \! \! \int\displaylimits_{D^+(\Sigma)} \! \! \rho(x') G(x,x') d V' 
    \nonumber
    \\
    &~ + \frac{1}{4 \pi} \int\displaylimits_\Sigma \left[ \psi(x') \nabla^{c'} G(x,x') - G(x,x') \nabla^{c'} \psi(x') \right] d S_{c'}.
    \label{KirchhoffScalar}
\end{align}
Here, $D^+(\Sigma)$ denotes the future domain of dependence of $\Sigma$. Tails exist when perturbations to the source or initial data can travel along timelike as well as null geodesics. It follows from \eqref{KirchhoffScalar} that this occurs when the support of $G$ includes regions where its arguments are timelike-separated. If no such regions exist, $G$ is said to be tail-free and $\psi$ satisfies Huygens' principle.

It has long been known that the retarded Green function associated with \eqref{ScalarField} in a plane wave background is tail-free \cite{Guenther}. In full, it is
\begin{equation}
    G(x,x') = \Delta^{1/2}(x,x') \delta_- (\sigma(x,x')).
    \label{ScalarG}
\end{equation}
$\delta_-( \sigma ) = \delta(\sigma) \Theta(x > x')$ and $\Theta(x > x')$ is a distribution which is equal to one if $x$ is in the future of $x'$ and vanishes otherwise. $\sigma$ denotes Synge's world function and $\Delta$ the van Vleck determinant. These two-point scalars are explicitly given by \eqref{Sigma} and \eqref{VanVleck} in plane wave spacetimes. Noting that $\Delta_{\mathrm{flat}} = 1$ everywhere in flat spacetime, the form of \eqref{ScalarG} is identical to that of the retarded Green function $G_\mathrm{flat} = \delta_- (\sigma_\mathrm{flat})$ associated with a flat background. Both solutions are concentrated only on light cones (where $\sigma =0$ or $\sigma_{\mathrm{flat}}=0$). The only qualitative difference between the flat and plane wave Green functions is that $\Delta$ is not constant in the plane wave case. This is essentially a geometric optics effect related to the focusing of null geodesics by the spacetime curvature.

In generic (not necessarily plane wave) spacetimes, the first term in \eqref{ScalarG} remains as-is. It describes the propagation of disturbances along null geodesics. A second term may also arise, however:
\begin{equation}
    G = \Delta^{1/2} \delta_- (\sigma) + V \Theta_- (-\sigma).
    \label{ScalarGAnsatz}
\end{equation}
$V$ is relevant for timelike-separated points, and is therefore referred to as the tail of $G$. If $V \neq 0$, disturbances propagate in timelike as well as null directions. The effects of localized perturbations may then persist long after they're first observed. Mathematically, it is straightforward to show that tails do not occur for massless scalar fields on plane wave backgrounds. Substituting \eqref{ScalarGAnsatz} into \eqref{GScalar} shows that in any spacetime,
\begin{equation}
    \Box V = 0
    \label{VBulkScalar}
\end{equation}
everywhere and
\begin{equation}
    \sigma^a \nabla_a V + \frac{1}{2} (\Box \sigma - 2) V = \frac{1}{2} \Box \Delta^{1/2}
    \label{VodeScalar}
\end{equation}
when $\sigma = 0$. The self-adjointness of \eqref{ScalarField} also implies that $V(x,x') = V(x',x)$. Fixing $x'$ and noting that $\sigma^a(x,x') = \nabla^a \sigma (x,x')$ is tangent to the geodesic connecting $x'$ to $x$, Eq. \eqref{VodeScalar} acts like an ordinary differential equation along each null geodesic which passes through $x'$. Integrating it provides characteristic initial data that can be used to solve \eqref{VBulkScalar} and obtain $V$ everywhere. In a plane wave background, recall from \eqref{VanVleck} that $\Delta^{1/2}(x,x')$ is a scalar field depending only on the phase coordinates $u,u'$ of its arguments. Any scalar of this type has vanishing d'Alembertian, so $\Box \Delta^{1/2} = 0$. The initial data for $V$ therefore vanishes and the unique solution to \eqref{VBulkScalar} is $V = 0$. This recovers \eqref{ScalarG}.

Tails typically do appear if the field equation \eqref{ScalarField} is modified in some way. Consider the addition of a mass term
\begin{equation}
    (\Box - \mu^2) \psi_\mu = - 4 \pi \rho.
\end{equation}
Solutions to this equation develop tails even in flat spacetime. Remarkably, the form of the tail in plane wave backgrounds is exactly the same as in the flat case. Using $J_1$ to denote a Bessel function of the first kind, the associated retarded Green function is
\begin{align}
    G_\mu &= \Delta^{1/2} \Big[ \delta_- (\sigma) 
    - \mu^2 \left( \frac{ J_1 ( \sqrt{-2 \sigma \mu^2}) }{ \sqrt{-2 \sigma \mu^2} } \right) \Theta_- (-\sigma) \Big].
    \label{GScalarMass}
\end{align}
Initially sharp perturbations to a massive scalar field do not appear sharp to timelike observers. Consider holding the second argument of $G_\mu(x,x')$ fixed and extending $x$ as far as possible into the future before encountering a $u=U$ wave surface where $\det B_{ij}(U,u') = 0$. Excluding degenerate cases where $B_{ij}(U,u')$ vanishes entirely, it is shown in \cite{HarteCaustics} that $\sigma \rightarrow - \infty$ and $(\Delta/|\sigma|)^{1/2}$ is finite for almost all approaches to such a hypersurface. The tail appearing in \eqref{GScalarMass} therefore tends to zero almost everywhere in this limit.

Of more direct physical interest is the behavior of an electromagnetic field $F_{ab}$. This is known to propagate without tails in plane wave backgrounds \cite{Kuenzle}. Unlike in the scalar case, gauge is an important consideration here. Green functions for vector potentials $A_a$ in commonly-used gauges do develop tails in plane wave backgrounds. These  can nevertheless be removed by appropriate gauge transformations $A_a \rightarrow A_a + \nabla_a \chi$, and are therefore physically irrelevant.

Consider a vector potential $A_a$ in Lorenz gauge. Introducing an electromagnetic current $J_a$, Maxwell's equations are equivalent to
\begin{equation}
    \Box A_a = - 4 \pi J_a, \qquad \nabla^a A_a = 0 
    \label{Maxwell}
\end{equation}
in vacuum spacetimes. The Hadamard ansatz for an associated retarded Green function is
\begin{equation}
    G_{a}{}^{a'} = U_{a}{}^{a'} \delta_- (\sigma) + V_{a}{}^{a'} \Theta_- (-\sigma) .
    \label{GreenEM}
\end{equation}
This must be a solution to
\begin{equation}
    \Box G_{a}{}^{a'} = - 4 \pi g_{a}{}^{a'} \delta
    \label{GreenEqEM}
\end{equation}
in a vacuum spacetime. As in the scalar case, the first (``direct'') term in $G_{a}{}^{a'}$ has the same form in any spacetime \cite{PoissonRev}:
\begin{equation}
    U_{a}{}^{a'} = \Delta^{1/2} g_{a}{}^{a'}.
    \label{UEM}
\end{equation}
$g_{a}{}^{a'}$ is the parallel propagator and $\Delta$ is again the van Vleck determinant. The appearance of $g_{a}{}^{a'}$ here may be interpreted as a reason why polarizations associated with electromagnetic waves are parallel-propagated in the limit of geometric optics. In plane wave backgrounds, $g_{a}{}^{a'}$ is explicitly given by \eqref{ParallelProp}.

The self-adjointness of the wave equation implies that the tail is symmetric in its arguments:
\begin{equation}
    V_{aa'} (x,x') = V_{a'a} (x,x').
    \label{RecipEM}
\end{equation}
Substituting \eqref{GreenEM} into \eqref{GreenEqEM} also shows that
\begin{equation}
    \Box V_{a}{}^{a'}  = 0 
    \label{VBulkEM}
\end{equation}
everywhere and
\begin{equation}
    \sigma^b \nabla_b V_{a}{}^{a'} + \frac{1}{2} (\Box \sigma - 2) V_{a}{}^{a'} = \frac{1}{2} \Box U_{a}{}^{a'} 
    \label{VodeEM}
\end{equation}
when $\sigma = 0$. Eq. \eqref{VodeEM} provides characteristic initial data for \eqref{VBulkEM}. Unlike in the massless scalar case, this data is not trivial. It may be found by first noting that the van Vleck determinant satisfies the transport equation \cite{PoissonRev}
\begin{equation}
    \sigma^a \nabla_a \Delta^{1/2} + \frac{1}{2}( \Box \sigma -4) \Delta^{1/2} = 0.
\end{equation}
Factoring this out of the tail by defining $W_{a}{}^{a'} = \Delta^{-1/2} V_{a}{}^{a'}$, \eqref{VodeEM} reduces to
\begin{align}
    \sigma^b \nabla_b W_{a}{}^{a'} + W_{a}{}^{a'} = \frac{1}{2} \Box g_{a}{}^{a'}.
    \label{VodeEM2}
\end{align}
The main advantage of \eqref{VodeEM2} over \eqref{VodeEM} is that the left-hand side of \eqref{VodeEM2} is a total derivative. Consider a null geodesic $y(u)$ such that $y(u') = x'$. Then,
\begin{equation}
    \frac{ D }{ d \tau } \left[ (\tau-u') W_{a}{}^{a'} (y(\tau),x') \right] = \frac{1}{2} \Box g_{a}{}^{a'} (y(\tau),x')
    \label{VodeEM3}
\end{equation}
on a plane wave background. Initial data for the tail is entirely determined by the degree to which parallel-transported vectors satisfy the vector wave equation. 

Directly computing the right-hand side of \eqref{VodeEM3} using \eqref{ParallelProp} and properties of the null tetrad $(\ell^a, n^a, m^a, \bar{m}^a)$ described in Sec. \ref{Sect:PlaneWaves}, 
\begin{equation}
    \Box g_{a}{}^{a'} = - \nabla^2 (\zeta \bar{\zeta} ) \ell_a \ell^{a'}. 
    \label{BoxParallel}
\end{equation}
Recalling that $\zeta$ is linear in the transverse coordinates $x^i, x'^i$ and is independent of $v,v'$, $\nabla^2 (\zeta \bar{\zeta})$ can depend only on the phase coordinates $u,u'$. The initial data for $V_{a}{}^{a'}$ is therefore
\begin{equation}
    V_{a}{}^{a'}(x,x') = -  \Delta^{1/2}(x,x') Z(u,u') \ell_a \ell^{a'},
    \label{TailEM}
\end{equation}
where
\begin{equation}
    (u-u') Z(u,u') =  \frac{1}{2} \int\displaylimits_{u'}^{u} d \tau \nabla^2 (\zeta \bar{\zeta})|_{(y(\tau),x')}.
    \label{ZDef}
\end{equation}
The reciprocity relation \eqref{RecipEM} implies that $Z(u,u') = Z(u', u)$.

Eq. \eqref{TailEM} arose as the solution to \eqref{VodeEM}. As such, it is \textit{a priori} valid only when its arguments are null-separated. Note, however, that \eqref{TailEM} is a solution to the bulk wave equation \eqref{VBulkEM} everywhere. No additional work is required to extend the tail into the interiors of null cones. The complete retarded Green function for Lorenz-gauge vector potentials in plane wave backgrounds is
\begin{align}
        G_{a}{}^{a'} = \Delta^{1/2} [ g_{a}{}^{a'} \delta_- (\sigma) - Z \ell_a \ell^{a'} \Theta_- (-\sigma) ].
        \label{GEM}
\end{align}

A general vector potential in Lorenz gauge may be derived from this Green function using the Kirchhoff integral
\begin{align}
    A_{a} &= \! \! \! \int\displaylimits_{D^+(\Sigma)} \! \! G_{b}{}^{a'} J_{a'}  d V' 
    \nonumber
    \\
    & ~+ \frac{1}{4 \pi} \int\displaylimits_\Sigma \Big( A_{a'} \nabla^{c'} G_{a}{}^{a'}   - G_{a}{}^{a'} \nabla^{c'} A_{a'} ) d S_{c'}.
    \label{AKirchhoff}
\end{align}
This produces solutions to the wave equation $\Box A_a = - 4\pi J_a$ in terms of the current $J_a$ and initial data prescribed on $\Sigma$. It does not enforce the gauge condition, however. Electromagnetic fields $F_{ab} = 2 \nabla_{[a} A_{b]}$ determined by \eqref{AKirchhoff} are true solutions to Maxwell's equations only if the initial data is chosen appropriately and the current is everywhere conserved.

It is evident from \eqref{GEM} and \eqref{AKirchhoff} that perturbations in $A_a$ generically propagate along timelike as well as null geodesics. The same cannot be said for perturbations to $F_{ab}$. This can be shown by applying an exterior derivative to \eqref{AKirchhoff} in order to obtain a Kirchhoff formula for $F_{ab}$. The relevant propagator is clearly $G_{ab}{}^{a'} = \nabla_{[a} G_{b]}{}^{a'}$, which has no tail:
\begin{equation}
    \nabla_{[a} V_{b]}{}^{a'} \Theta_-(-\sigma)= 0 . 
    \label{NoTail}
\end{equation}
$F_{ab}$ therefore satisfies Huygens' principle on plane wave backgrounds, as claimed.


%
%

\section{Gravitational perturbations}
\label{Sect:GravGreen}

At least within normal neighborhoods, both scalar and electromagnetic test fields on plane wave backgrounds behave essentially as they do in flat spacetime: Pulses of radiation which are initially sharp remain sharp as they propagate. We now consider perturbations to the plane wave metric itself. 

Metric perturbations on plane wave spacetimes have previously been considered in several contexts. The theory of colliding plane waves is particularly well-developed, containing many exact solutions to the fully nonlinear Einstein equation \cite{CollidingPlane}. Yurtsever has also discussed colliding waves which are planar only over large but finite regions \cite{Yurtsever, Yurtsever2}.

These results do not allow for localized perturbations and are therefore inadequate to discuss the presence of tails. Linear perturbations of plane wave spacetimes which are not plane-symmetric have been derived before \cite{XanthPlane, XanthPlane2}, although not in a form which appears to be useful for our purposes. The approach taken here is to instead describe perturbations of plane wave spacetimes in terms of a Green function associated with the linearized Einstein equation. 

As is usual in general relativistic perturbation theory, we consider a 1-parameter family of metrics
\begin{equation}
    \hat{g}_{ab}(\epsilon) = g_{ab} + \epsilon h_{ab} + O(\epsilon^2)
\end{equation}
which tend smoothly to a vacuum plane wave background $g_{ab}$ as $\epsilon \rightarrow 0^+$. Only first order perturbations are considered here. Carets are used to denote quantities associated with the full (background + perturbation) spacetime. All indices are raised and lowered using the background metric $g_{ab}$. 

It is convenient to work in the Lorenz gauge. This is defined by demanding that the  trace-reversed metric perturbation $h_{ab}^{\mathrm{TR}} = h_{ab} - g_{ab} g^{cd} h_{cd}/2$ be divergence-free:
\begin{equation}
    \nabla^b h_{ab}^{\mathrm{TR}} = 0.
    \label{Gauge}
\end{equation}
Einstein's equation linearized about $g_{ab}$ is then 
\begin{equation}
    \Box h^{\mathrm{TR}}_{ab} + 2 R_{a}{}^{c}{}_{b}{}^{d} h^{\mathrm{TR}}_{cd} = - 16 \pi T_{ab},
    \label{EinstLin}
\end{equation}
where $T_{ab}$ denotes the perturbed stress-energy tensor. All derivative operators are those associated with the background.

General solutions to \eqref{EinstLin} may be constructed using Green functions. Consider a two-point distribution $G^{ab}{}_{a'b'}(x,x')$ which is separately symmetric in the indices $ab$ and $a'b'$ and which satisfies 
\begin{equation}
    \Box G_{ab}{}^{a'b'} + 2 R_{a}{}^{c}{}_{b}{}^d G_{cd}{}^{a'b'} = - 4 \pi g_{a}{}^{(a'} g_{b}{}^{b')} \delta. 
    \label{GravGreenPDE}
\end{equation}
The Hadamard ansatz for a retarded solution to this equation is
\begin{equation}
    G_{ab}{}^{a'b'} = U_{ab}{}^{a'b'} \delta_- (\sigma) + V_{ab}{}^{a'b'} \Theta_- (-\sigma).
    \label{HadamardGrav}
\end{equation}
In every spacetime, the first term here is known to be given by \cite{PoissonRev}
\begin{equation}
    U_{ab}{}^{a'b'} = g_{a}{}^{(a'} g_{b}{}^{b')} \Delta^{1/2}.
    \label{UGrav}
\end{equation}
The polarization tensor of a gravitational wave is therefore parallel-propagated in the limit of geometric optics. Its ``amplitude'' also varies according to the usual focusing factor $\Delta^{1/2}$. 

The gravitational tail $V_{ab}{}^{a'b'}$ appearing in \eqref{HadamardGrav} does not have a universal form. It satisfies
\begin{equation}
    \Box V_{ab}{}^{a'b'} + 2 R_{a}{}^{c}{}_{b}{}^{d} V_{cd}{}^{a'b'} = 0
    \label{VBulk}
\end{equation}
everywhere with characteristic initial data determined by
\begin{align}
    \sigma^c \nabla_c V_{ab}{}^{a'b'} + \frac{1}{2} (\Box \sigma - 2) V_{ab}{}^{a'b'} = \frac{1}{2}  \Box U_{ab}{}^{a'b} 
    \nonumber
    \\
    ~  + R_{c}{}^{a}{}_{d}{}^b U_{ab}{}^{a'b'}
    \label{Vode}
\end{align}
for all null-separated points $x,x'$. The self-adjointness of \eqref{EinstLin} implies the reciprocity relation
\begin{equation}
    V_{aba'b'} (x,x') = V_{a'b'ab}(x',x). 
    \label{RecipGrav}
\end{equation}

As in the scalar and electromagnetic cases discussed above, $G_{ab}{}^{a'b'}$ may be used to relate general solutions of the relaxed Einstein equation \eqref{EinstLin} to sources and initial data using the Kirchhoff integral
\begin{align}
    &h^{\mathrm{TR}}_{ab} = 4 \!\! \int\displaylimits_{D^+(\Sigma)} \! \! G_{ab}{}^{a'b'} T_{a'b'} d V'
    \nonumber
    \\
    & ~ + \frac{1}{4\pi} \int\displaylimits_{\Sigma}   \! \left( \nabla^{c'} G_{ab}{}^{a'b'} h^{\mathrm{TR}}_{a'b'}  - G_{ab}{}^{a'b'} \nabla^{c'} h^{\mathrm{TR}}_{a'b'} \right) dS_{c'}.
    \label{KirchhoffGrav}
\end{align}
Only perturbations which satisfy the gauge condition \eqref{Gauge} are true solutions to the linearized Einstein equation. As a consequence, $\nabla^a T_{ab} = 0$ and the initial data appearing in \eqref{KirchhoffGrav} must satisfy the gauge condition in an appropriate sense.



%
%

One component of $V_{ab}{}^{a'b'}$ may be derived immediately. Taking the trace of \eqref{EinstLin} and noting the appropriate boundary conditions, the gravitational tail $V_{ab}{}^{a'b'}$ is related to the tail $V$ associated with massless scalar fields via $g^{ab} V_{ab}{}^{a'b'} = g^{a'b'} V$ \cite{PoissonRev}. But $V = 0$ in plane wave backgrounds, so
\begin{equation}
    g^{ab} V_{aba'b'} = V_{aba'b'} g^{a'b'} = 0. 
    \label{TailTrace}
\end{equation}
$g^{ab} h_{ab}$ therefore propagates only along null geodesics in Lorenz gauge. This is the sense in which Waylen \cite{Waylen} has claimed that metric perturbations on plane wave backgrounds satisfy Huygens' principle. Other components of $h_{ab}$ do not behave so simply. To see this, it is sufficient to note that regularity of $V_{ab}{}^{a'b'}$ in \eqref{Vode} implies that the tail has the coincidence limit \cite{PoissonRev}
\begin{equation}
    [V_{abc'd'} ] = R_{ac'bd'}.
    \label{Vcoinc}
\end{equation}
The right-hand side of this equation doesn't vanish, so $V_{ab}{}^{a'b'}$ cannot vanish either.

As in the electromagnetic case above, it is convenient to factor the van Vleck determinant out of the tail by defining
\begin{equation}
    W_{ab}{}^{a'b'} = \Delta^{-1/2} V_{ab}{}^{a'b'}.
    \label{WDef}
\end{equation}
Eq. \eqref{Vode} then reduces to
\begin{align}
    \frac{D}{d\tau} \left[ (\tau-u') W_{ab}{}^{a'b'} \right] = \frac{1}{2} \Box (g_{a}{}^{(a'} g_{b}{}^{b')}) 
\nonumber
\\
~+ R^{c}{}_{a}{}^{d}{}_{b} g_{c}{}^{(a'} g_{d}{}^{b')}
    \label{Vode2}
\end{align}
along null geodesics passing through $x'$. The characteristic initial data needed to obtain the tail follows from this equation by using \eqref{ParallelProp} and contracting with all possible tetrad components. The resulting calculation is tedious but straightforward. It results in an expression for $V_{ab}{}^{a'b'}$ which is valid when $\sigma=0$. Whatever is not taken into account by this data may be parametrized by the addition of $\sigma L_{ab}{}^{a'b'}$ for some smooth $L_{ab}{}^{a'b'}$. The result of these computations is that
\begin{align}
    V_{ab}{}^{a'b'} = \Delta^{1/2} \Re \Big\{ \sigma L_{ab}{}^{a'b'} +\big( \bar{m}_a \bar{m}_b Y + 4 Z m_{(a} \bar{m}_{b)} \big) \ell^{a'} \ell^{b'} 
    \nonumber
    \\
    ~ + \ell_a \ell_b \big[ g_{c}{}^{a'} g_{d}{}^{b'} \big( Y \bar{m}^c \bar{m}^d + 4 Z m^{(c} \bar{m}^{d)} \big) - Z g^{a'b'} \big] 
    \nonumber
    \\
    ~  - 2 \ell_{(a} m_{b)} \ell^{(a'} g_{c}{}^{b')} ( \bar{Y} m^c + 4 Z \bar{m}^c) - Z g_{ab} \ell^{a'} \ell^{b'} \Big\} . 
    \label{TailGrav}
\end{align}
The real function $Z(u,u')$ appearing here also arose in the electromagnetic Green function $G_{a}{}^{a'}$ and is defined by \eqref{ZDef}. $Y(u,u') = Y(u',u)$ is complex and satisfies
\begin{equation}
     (u-u') Y(u,u') =  \int\displaylimits_{u'}^{u} d \tau \left( \nabla^2 ( \zeta^2 ) - 2 \mathcal{H}  \right).
     \label{YDef}
\end{equation}

The only portion of the gravitational Green function which is left to be determined is $L_{ab}{}^{a'b'}$. Substituting \eqref{TailGrav} into \eqref{VBulk} while using \eqref{ZDef} and \eqref{YDef} shows that
\begin{equation}
    L_{ab}{}^{a'b'} = - \frac{1}{4} (|Y|^2 + 8 Z^2) \ell_a \ell_b \ell^{a'} \ell^{b'}.
    \label{LDef}
\end{equation}
Note that the scalar coefficient in this equation depends only on $u$ and $u'$. Together, \eqref{ZDef}, \eqref{HadamardGrav}, \eqref{UGrav}, and \eqref{TailGrav}-\eqref{LDef} completely describe the retarded Green function for Lorenz-gauge metric perturbations on vacuum plane wave backgrounds. This appears to be the first closed-form example of a Green function for Einstein's equation linearized off of a non-flat vacuum background (see, however, \cite{CosmGreen, CosmGreen2} for discussions of Green functions on Friedmann-Robertson-Walker backgrounds).

The scalar, electromagnetic, and gravitational Green functions derived here all have the remarkable property that their Hadamard series terminate at finite order. In generic spacetimes, tails of Green functions can be obtained systematically using the series expansion
\begin{equation}
    V^{\cdots} (x,x') = \sum_{n=0}^\infty V^{\cdots}_n(x,x') \sigma^n(x,x').
    \label{HadSeries}
\end{equation}
This is called a Hadamard series, and is motivated by the observation that initial data equations such as \eqref{Vode} are ordinary differential equations which apply when $\sigma = 0$ (see, e.g., \cite{Friedlander}). Terms which are higher order in $\sigma$ effectively expand the solution ``into the light cones.'' Conditions may be placed on the coefficients $V^{\cdots}_n$ such that these functions are uniquely determined by ordinary differential equations for all $n$. The results of Sec. \ref{Sect:EM} imply that all Hadamard coefficients vanish for massless scalar fields propagating on plane wave backgrounds. For Lorenz-gauge vector potentials, $V_n^{aa'}$ vanishes for all $n \geq 1$. The tail associated with Lorenz-gauge metric perturbations admits non-vanishing Hadamard coefficients only when $n=0$ or $n=1$. In all of these cases, Hadamard expansions result in finite series. More precisely, the Hadamard coefficients $V^{\cdots}_n$ vanish for all $n \geq (\mbox{tensor rank of the relevant field})$. It is not clear if this pattern continues for higher-rank fields propagating on plane wave backgrounds.

\subsection*{Plane-fronted perturbations}

The tail associated with $G_{ab}{}^{a'b'}$ is discussed in detail in Sec. \ref{Sect:PlaneToPlane}. One interesting class of perturbations may, however, be understood immediately. Consider sources with the form $T_{ab} = \mathfrak{T} \ell_a \ell_b$ and initial data satisfying
\begin{equation}
    h_{ab} \propto \ell_a \ell_b, \qquad d S_c \nabla^c h_{ab} \propto \ell_a \ell_b
\end{equation}
on a spacelike hypersurface $\Sigma$. Eqs. \eqref{ParallelProp}, \eqref{HadamardGrav}, \eqref{UGrav}, \eqref{KirchhoffGrav}, \eqref{TailGrav}, and \eqref{LDef} imply that the tail cannot affect any metric perturbation to the future of $\Sigma$. The relevant portion of the Green function is simply
\begin{equation}
    G_{ab}{}^{a'b'} \ell_{a'} \ell_{b'} = \Delta^{1/2} \delta_-(\sigma) \ell_a \ell_b,
\end{equation}
from which it follows that
\begin{equation}
    h_{ab} = \delta H \ell_a \ell_b
    \label{ppPert}
\end{equation}
for some scalar field $\delta H$. Stress-energy conservation and the Lorenz gauge condition require
\begin{equation}
    \mathcal{L}_\ell \mathfrak{T} = \partial_v \mathfrak{T} = \partial_v \delta H = 0.
\end{equation}
Perturbations in this class therefore share some degree of plane symmetry with the background spacetime. 

More than this, the perturbed metric is exactly that of a (not necessarily vacuum) pp-wave spacetime with the profile $H_{ij}(u) x^i x^j + \epsilon \delta H(u,x^k)$. Although this result has been obtained using the linearized Einstein equation, it is actually an exact solution to the fully nonlinear Einstein equation. That the linearized solution is also an exact solution is a consequence of the fact that the perturbation \eqref{ppPert} is in extended Kerr-Schild form \cite{Xanth}.

\section{Interpreting the tail}
\label{Sect:PlaneToPlane}

It was shown in Sec. \ref{Sect:EM} that tails associated with Lorenz-gauge vector potentials are pure gauge in plane wave backgrounds. It is interesting to ask whether a similar situation arises for the tail associated with the gravitational Green function $G_{ab}{}^{a'b'}$ derived in Sec. \ref{Sect:GravGreen}. In the electromagnetic case, all physical information is contained in the Faraday tensor $F_{ab}$. A ``relevant'' electromagnetic tail therefore exists if and only if initial data for $F_{ab}$ propagates along timelike curves. It follows from \eqref{NoTail} that this does not occur, so any tails associated with the vector potential are unphysical.

Unfortunately, there is no direct analog of $F_{ab}$ which can describe the geometry of spacetime in a way that is both complete and gauge-invariant. The perturbed metric changes drastically in different gauges, and is therefore difficult to interpret directly. The perturbed Riemann tensor is less dependent on gauge, but not completely independent of it. Nevertheless, the curvature may be used to construct various quantities which \textit{are} gauge-invariant. One possible route is to compute perturbations to all scalars  
\begin{equation}
    R^{abcd} R_{abcd}, \quad \nabla^a R^{bcdf} \nabla_a R_{bcdf}, \quad \ldots 
\end{equation}
formed from local polynomials in the Riemann tensor and its derivatives (together with $g_{ab}$ and $\epsilon_{abcd}$). Every scalar of this type is known to vanish in the background spacetime. Their perturbations are therefore gauge-invariant. Local knowledge of all curvature scalars places strong constraints on the local geometry \cite{CurvScalars}. Here, all tail-related perturbations to the curvature scalars may be shown to vanish. This is a gauge-invariant statement. It does not, however, exhaust all physically-relevant information which may be extracted from $h_{ab}$.

Geometries with vanishing curvature scalars are known to comprise a certain subset of the Kundt spacetimes \cite{VSI}. A metric $\hat{g}_{ab}$ is said to be of the Kundt type if it admits a null vector field $\hat{\ell}^a$ which is geodesic, expansion-free, shear-free, and twist-free. All pp-waves (and therefore plane waves) are in this class. Our strategy to understand the tail portion of $G_{ab}{}^{a'b'}$ is to first identify a perturbation $\hat{\ell}^a$ of the background vector field $\ell^a$ which establishes the perturbed spacetime to locally be in the Kundt class. Indeed, we obtain a nontrivial null vector field whose first derivative vanishes entirely: $\hat{\nabla}_b \hat{\ell}^a = O(\epsilon^2)$. This implies that through first order in the small parameter $\epsilon$, tail effects can only perturb a background plane wave into a pp-wave. We construct an explicit gauge transformation which transforms the perturbed metric $\hat{g}_{ab}$ into the canonical Brinkmann form \eqref{ppMetric}. The perturbed wave profile may then be read off directly from the metric components. It is found to be at most quadratic in the transverse coordinates, so the perturbed spacetime is locally a plane wave and not a more general type of pp-wave. 

Before carrying out this procedure, it is first important to note that no universal properties can be expected to hold for the perturbed metric as a whole and for completely arbitrary initial data. Only the ``tail components'' of metric perturbations might be expected to have a simple description for all possible choices of initial data. This concept must be made precise. In general, it is difficult to disentangle a meaningful tail from the remainder of a generic metric perturbation. Naively, the Kirchhoff integral \eqref{KirchhoffGrav} could be used to isolate only those portions of $h_{ab}(x)$  which involve data in the interior of the past light cone of $x$. This could be called the ``tail'' $h^\flat_{ab}(x)$ of the full metric perturbation at a particular point $x$. It is, however, impossible to make geometrically-meaningful statements knowing only a metric at a single point. Derivatives of $h^\flat_{ab}$ must also be computed at $x$. These have Kirchhoff representations of their own which involve data that is no longer guaranteed to be confined only to the interior of the past light cone of $x$. Derivatives of objects which appear to be ``pure tails'' are not pure tails themselves. This is because initial data which lies sufficiently close to -- but not on -- the past light cone of $x$ can intersect the past light cone of a neighboring point $x+ dx$. 

Although our assumptions can be weakened, this problem is avoided here by considering only those cases where it does not arise. First suppose for simplicity that $T_{ab} = 0$. Fixing a particular point $x$, let all nontrivial data on the initial hypersurface $\Sigma$ vanish where that hypersurface meets the past light cone of $x$. Also demand that similar conditions exist for all points in an open neighborhood of $x$. This corresponds to considering general vacuum perturbations observed only in regions which are timelike- or spacelike- (but not null-) separated to all nontrivial initial data. 

Under these restrictions, various properties of $h_{ab}$ may be deduced directly from inspection of \eqref{KirchhoffGrav} and \eqref{TailGrav}. The first of these is that $g^{ab} h_{ab} = 0$. Using the notation $h_{n\ell} = h_{ab} n^a \ell^b$, it follows that $h_{n\ell} = h_{m \bar{m}}$. Additionally, $h_{\ell \ell} = h_{m \ell} =  0$. The components $h_{mm}$, $h_{m \bar{m}}$, and $h_{\ell n}$ depend only on the phase coordinate $u$. $h_{mn}$ may depend on both $u$ and $x^i$, and is at most linear in the latter coordinates. $h_{nn}$ is at most linear in $v$ (with a coefficient depending only on $u$) and at most quadratic in $x^i$. Together, these observations strongly constrain the geometric character of the perturbed spacetime. 

Recall that the vector field $\ell^a = \partial/\partial v$ is both null and covariantly constant with respect to the background metric $g_{ab}$. $\ell^a$ is also null with respect to the perturbed metric $\hat{g}_{ab}$. It is easily verified, however, that $\hat{\nabla}_b \ell^a \neq 0$ in general. Nevertheless, the rescaled vector field
\begin{equation}
    \hat{\ell}^a = ( 1 + \epsilon \alpha) \ell^a
    \label{EllHat}
\end{equation}
is both null and constant with respect to $\hat{g}_{ab}$ [through $O(\epsilon)$] if the real scalar field $\alpha(u)$ satisfies
\begin{equation}
    \dot{\alpha} = \frac{1}{2} \partial_v h_{nn}. 
    \label{alphaDef}
\end{equation}
That such an $\hat{\ell}^a$ exists is highly nontrivial. It implies that $\hat{g}_{ab}$ must describe a pp-wave through $O(\epsilon)$. The lack of uniqueness associated with $\alpha$ corresponds to the usual freedom to rescale $\hat{\ell}^a$ by a multiplicative constant. 

Any pp-wave spacetime must admit not only a constant null vector field, but also a constant null 2-form. In the background spacetime, examples of such 2-forms are the $f_{ab}$ and $f^*_{ab}$ of Eq. \eqref{fDef}. The complex 2-form $\ell_{[a} m_{b]} = (f_{ab} - i f_{ab}^*)/\sqrt{2}$ and its conjugate are null and constant as well. Preferred perturbations $\hat{m}^a$ to the background tetrad component $m^a$ may therefore be found by demanding that $\hat{\ell}^{[a} \hat{m}^{b]}$ be null and constant with respect to $\hat{g}_{ab}$ (together with the usual requirements that $\hat{m}^a$ be null, orthogonal to $\hat{\ell}^a$, and that $\hat{g}_{ab} \hat{m}^a \bar{\hat{m}}^b = 1$). This results in
\begin{align}
    \hat{m}^a = [1 + \epsilon (i \theta- \frac{1}{2} h_{m \bar{m}}) ] m^a - \frac{1}{2} \epsilon h_{mm} \bar{m}^a 
    \nonumber
    \\
    ~  - \epsilon \partial_{\bar{\xi}} \gamma \ell^a.
    \label{mPert}
\end{align}
$\hat{m}^a$ is considerably less unique than $\hat{\ell}^a$. It depends on the choice of an arbitrary real function\footnote{Only $\partial_{\bar{\xi}} \gamma$ is relevant for $\hat{m}^a$. $\gamma$ appears undifferentiated in the coordinate transformation \eqref{vHat}.} $\gamma(u, \xi, \bar{\xi})$. $\hat{m}^a$ also depends on $\theta(u)$, which is real and satisfies
\begin{equation}
    \dot{\theta} = \Im (\partial_{\xi} h_{mn}).
    \label{betaDef}
\end{equation}
Here, $\Im$ denotes the imaginary component of its argument. Different solutions for $\theta$ correspond to the usual freedom to vary $\hat{m}^a$ by a constant complex phase.

Given particular choices for $\hat{\ell}^a$ and $\hat{m}^a$ within the class just described, a unique real vector field $\hat{n}^a$ may be constructed which is null, satisfies $\hat{g}_{ab} \hat{\ell}^a \hat{n}^b = -1$, and is orthogonal to $\hat{m}^a$ and $\bar{\hat{m}}^a$:
\begin{align}
    \hat{n}^a = [ 1 + \epsilon (h_{m \bar{m}} -\alpha) ] n^a + \frac{1}{2} \epsilon h_{nn} \ell^a 
    \nonumber
    \\
    ~ - 2 \epsilon \Re \left[ (\partial_{\bar{\xi}} \gamma + h_{m n } )  \bar{m}^a \right].
    \label{nPert}
\end{align}
Together, the vector fields $(\hat{\ell}^a, \hat{n}^a, \hat{m}^a, \bar{\hat{m}}^a)$ form a null tetrad in the perturbed spacetime with very similar properties to the null tetrad $(\ell^a, n^a, m^a, \bar{m}^a)$ defined by \eqref{ellDef} and \eqref{TetradDef} in the background spacetime. The perturbed tetrad is not unique. It depends on a real function $\gamma(u, \xi, \bar{\xi})$ as well as two real numbers corresponding to initial conditions for the differential equations \eqref{alphaDef} and \eqref{betaDef}. 

Given that $\hat{\ell}^a$ is constant, the 1-form $\hat{g}_{ab} \hat{\ell}^b$ must be closed. It follows that there exists a scalar field $\hat{u}$ such that
\begin{equation}
    \hat{g}_{ab} \hat{\ell}^b = - \nabla_a \hat{u}.
    \label{uHatDef}
\end{equation}
By analogy with \eqref{uDef}, this defines a phase coordinate $\hat{u}$ for the perturbed spacetime. Similarly, scalar fields $\hat{v}$ and $\hat{\xi}$ may be introduced such that
\begin{equation}
    \hat{\ell}^a = \frac{\partial}{\partial \hat{v}}, \qquad \bar{\hat{m}}^a = \frac{\partial}{\partial \hat{\xi}}.
    \label{NewTetrad1}
\end{equation}
Coordinates with these properties fix seven out of ten metric components. The remainder follow from demanding that
\begin{equation}
    \hat{n}^a = \frac{\partial}{\partial \hat{u} } + \mathfrak{h} \frac{\partial}{\partial \hat{v}}
    \label{NewTetrad2}
\end{equation}
for some real $\mathfrak{h}(\hat{u}, \hat{\xi}, \bar{\hat{\xi}})$ [cf. \eqref{TetradDef}]. The perturbed metric then reduces to the explicit Brinkmann form
\begin{equation}
    \widehat{d s}^2 = 2 \left( -d \hat{u} d \hat{v} + \mathfrak{h} d\hat{u}^2 + d \hat{\xi} d \bar{\hat{\xi}} \right) + O(\epsilon^2)
    \label{PertMetric}
\end{equation}
for a pp-wave with profile $2 \mathfrak{h}$. 

Coordinate (or gauge) transformations $(u, v, \xi, \bar{\xi}) \rightarrow (\hat{u}, \hat{v}, \hat{\xi}, \bar{\hat{\xi}})$ with these properties may be constructed explicitly and used to write $\mathfrak{h}$ in terms of $h_{ab}$. Noting \eqref{EllHat} and \eqref{uHatDef}, the perturbed and background phase coordinates are related by
\begin{equation}
    \hat{u}(u) = u + \epsilon \beta(u), \qquad \dot{\beta}(u) = \alpha(u) - h_{m \bar{m}}(u).
    \label{lambdaDef}
\end{equation}
Different choices for $\alpha$ and $\beta$ correspond to the fact that a constant affine transformation applied to any valid phase coordinate produces another valid phase coordinate. The remaining coordinate transformations which imply \eqref{NewTetrad1} and \eqref{NewTetrad2} are 
\begin{align}
    \hat{v}(u,v,\xi,\bar{\xi}) = \big[ 1- \epsilon \alpha(u)\big] v + \epsilon \gamma (u,\xi, \bar{\xi}),
    \label{vHat}
    \\
    \hat{\xi}(u,\xi,\bar{\xi}) = \big[ 1 + \epsilon \big( i \theta(u) + \frac{1}{2} h_{m \bar{m}}(u)  \big) \big] \xi 
    \nonumber
    \\
    ~ + \frac{1}{2} \epsilon h_{mm}(u) \bar{\xi} + \epsilon \lambda (u) , 
\end{align}
where $\lambda(u)$ is complex and satisfies
\begin{align}
    \dot{\lambda} 
    =  \partial_{\bar{\xi}} \gamma + \big[ h_{mn} - i \xi \Im (\partial_{\xi} h_{mn} )\big]
    \nonumber
    \\
    ~ - \frac{1}{2}  \partial_u (h_{m \bar{m}} \xi + h_{mm} \bar{\xi})  .
    \label{muConstraint}
\end{align}
$\lambda$ can exist only if the right-hand side of this last equation is independent of both $\xi$ and $\bar{\xi}$. As a consequence, $\gamma$ is constrained by the two integrability conditions
\begin{align}
    \partial_\xi \partial_{\bar{\xi}} \gamma = \frac{1}{2} \partial_u h_{m \bar{m}} - \Re (\partial_{\xi} h_{mn}) ,
    \label{Int1}
    \\
    \partial_{\bar{\xi}}^2  \gamma = \frac{1}{2} \partial_u h_{m m} - \partial_{\bar{\xi}} h_{mn} .
    \label{Int2}
\end{align}
The right-hand sides of both of these relations depend only on $u$, so $\gamma$ must be quadratic in $\xi, \bar{\xi}$. For any real function $\gamma_0(u)$, the most general $\gamma$ is
\begin{align}
    \gamma = \gamma_0 + \Re\Big\{2 \big[\dot{\lambda} - (h_{mn}|_{\xi = 0} ) \big] \bar{\xi}+ \frac{1}{2} \xi \bar{\xi} \partial_v h_{nn}
    \nonumber
    \\
    ~ + \frac{1}{2} \bar{\xi}^2 (\partial_u h_{m m} - 2 \partial_{\bar{\xi}} h_{m n}  ) \Big\}.
    \label{gamma}
\end{align}
The Lorenz gauge condition \eqref{Gauge} has been used here in the form
\begin{equation}
    \Re (\partial_{\xi} h_{mn}) = \frac{1}{2} ( \partial_u h_{m \bar{m}} - \partial_v h_{nn} ) .
    \label{NewGauge}
\end{equation}
It is clear from \eqref{gamma} that the sub-quadratic components of $\gamma$ may be varied arbitrarily by appropriate choices of $\lambda$ and $\gamma_0$.

This freedom can be used to simplify the wave profile $\mathfrak{h}$. Applying the various coordinate transformations and comparing to \eqref{NewTetrad2},
\begin{align}
    \mathfrak{h} = \Re\Big\{ \Big[ \big( 1 + 2 \epsilon ( i \theta - \alpha) \big) \mathcal{H} - \epsilon \beta \dot{\mathcal{H}} \Big]  \bar{\hat{\xi}}^2 - \epsilon \mathcal{H} h_{\bar{m} \bar{m}} \hat{\xi} \bar{\hat{\xi}}
    \nonumber
    \\
    ~ - 2 \epsilon  \bar{\lambda} \mathcal{H} \bar{\hat{\xi}}
     + \frac{1}{2} ( h_{nn} - \hat{v} \partial_v h_{nn}) + \partial_u \gamma  \Big\}. 
    \label{hGen}
\end{align}
This has been written in terms of the perturbed Brinkmann coordinates $(\hat{u}, \hat{v}, \hat{\xi}, \bar{\hat{\xi}})$. It satisfies $\partial_{\hat{v}} \mathfrak{h} = 0$, as required for a general pp-wave profile. Furthermore, $\mathfrak{h}$ is at most quadratic in the transverse coordinates $\hat{\xi}, \bar{\hat{\xi}}$. This means that the perturbed spacetime is actually a plane wave and not another type of pp-wave.

$\hat{g}_{ab}$ satisfies the linearized Einstein equation by construction, so
\begin{equation}
    \partial_{\hat{\xi}} \partial_{\bar{\hat{\xi}} } \mathfrak{h} = 0.
    \label{hCrossTerms}
\end{equation}
Further simplifications arise from appropriate choices for $\gamma_0$ and $\lambda$. Letting
\begin{equation}
    \dot{\gamma}_0 = - \frac{1}{2} ( h_{nn} - v \partial_v h_{nn}) |_{\xi = 0} 
    \label{gamma0}
\end{equation}
eliminates all terms in $\mathfrak{h}$ which are independent of the transverse coordinates. Different solutions to \eqref{gamma0} correspond to definitions for $\hat{v}$ coordinates which differ by additive constants.

All terms linear in $\hat{\xi}$ and its conjugate which appear in $\mathfrak{h}$ may be eliminated by demanding that $\lambda$ satisfy
\begin{equation}
   \ddot{\lambda} = \mathcal{H} \bar{\lambda} + (\partial_u h_{mn}  - \frac{1}{2} \partial_{\bar{\xi}} h_{nn})_{\xi=0}. 
   \label{lambda}
\end{equation}
In the absence of metric perturbations, this reduces to the background geodesic equation \eqref{TransverseCoordsIm}. Different solutions for $\lambda$ affect the definitions for both $\hat{v}$ and $\hat{\xi}$ in ways which may be identified with motions generated by Killing fields. 

Applying \eqref{hCrossTerms}-\eqref{lambda} to \eqref{hGen} reduces the wave profile to the canonical form [cf. \eqref{PlaneWaveMetric}]
\begin{equation}
    \mathfrak{h}(\hat{u}, \hat{\xi}, \bar{\hat{\xi}}) =\Re \big[  \hat{\mathcal{H}}( \hat{u} ) \bar{\hat{\xi}}^2 \big],
\end{equation}
with the perturbed waveform
\begin{align}
    \hat{\mathcal{H}}   = & \big[ 1 + 2 \epsilon ( i \theta - \alpha) \big] \mathcal{H} - \epsilon \beta \dot{\mathcal{H}} 
    \nonumber
    \\
    &~ + \frac{1}{2} \epsilon ( \partial_{\bar{\xi}}^2 h_{nn} + \partial_u^2 h_{mm} -2  \partial_u \partial_{\bar{\xi}} h_{mn}) . 
    \label{WaveformPert}
\end{align}
Through $O(\epsilon)$, the tail of any linear perturbation to a plane wave spacetime with waveform $\mathcal{H}$ is another plane wave spacetime with waveform $\hat{\mathcal{H}}$. This is a complete description for the local geometry, and is unique up to the 3-parameter family of rescalings described by \eqref{WaveformGauge}. The integration constants associated with solving \eqref{alphaDef}, \eqref{betaDef}, and \eqref{lambdaDef} for $\alpha$, $\beta$, and $\theta$ represent infinitesimal versions of these rescalings and may therefore be ignored.

\begin{figure}[t]
    \centering
    \vskip .3 cm
    \includegraphics[width=.7\linewidth]{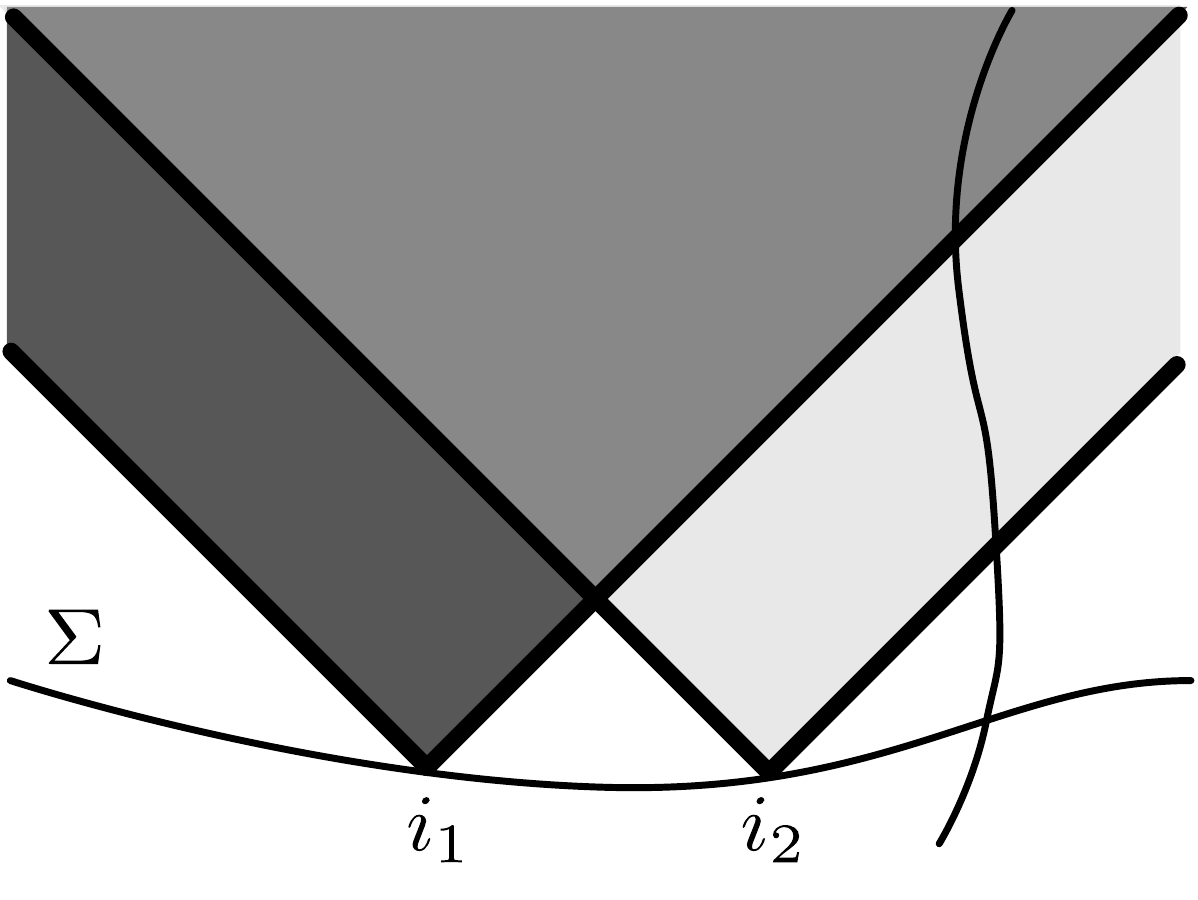}
    
    \caption{An initial value problem for vacuum perturbations on a plane wave spacetime. Nontrivial initial data is imposed on $\Sigma$ only in small regions around the two points $i_1, i_2$. The unshaded regions are locally determined by the background plane wave. The three shaded regions are locally plane waves, each of which may have different waveforms. The (finite-width) boundaries between each of the plane wave regions can be arbitrarily complicated. Also illustrated is a timelike worldline representing the path of a local observer.}
    \label{Fig:Cauchy}
\end{figure}

Using the Kirchhoff representation \eqref{KirchhoffGrav} for $h_{ab}$, the perturbed waveform can be written as an integral over initial data specified on a spacelike hypersurface $\Sigma$. Using the explicit form \eqref{TailGrav} for the tail $V_{ab}{}^{a'b'}$, $\hat{\mathcal{H}}$ may be shown to depend only on $ \hat{g}_{\ell \ell}|_\Sigma = \epsilon h_{\ell \ell}|_\Sigma$ and the first derivative of this quantity normal to $\Sigma$. In this sense, plane wave tails arise from only a single component of the initial metric perturbation.

\section{Discussion}
\label{Sect:Discussion}

At least within normal neighborhoods, we have shown that linear tails in vacuum general relativity cannot perturb a gravitational plane wave into anything other than another plane wave. The scattering of two gravitational waves -- one an initially-localized burst and the other a much stronger plane wave -- can leave behind only a very particular type of tail. This situation is illustrated in Fig. \ref{Fig:Cauchy} for the particular case where initial metric perturbations exist only in the vicinities of two distinct points $i_1$ and $i_2$. Regions which are null-separated from the initial data near $i_1$ and $i_2$ do not have any universal form. Everywhere else in the future of $\Sigma$ is locally a plane wave. Different plane wave regions are separated by the future light cones of the nontrivial initial data, and may all have different waveforms.

Except for two brief events associated with the wavefronts expanding outwards from $i_1$ and $i_2$, the timelike observer illustrated in Fig. \ref{Fig:Cauchy} can use purely local measurements to conclude that the surrounding geometry is always a plane wave. The instantaneous waveform of that plane wave can also be determined using local measurements. Such an observer could not, however, conclusively determine whether or not the waveform observed immediately after a particular wavefront is really ``different'' from the waveform observed before that wavefront. Such a determination requires the use of experiments which sample the geometry over finite distances.

These results are but one application of the Green function $G_{ab}{}^{a'b'}$ derived in Sec. \ref{Sect:GravGreen}. Another possible application concerns the effect of a compact object's ``own gravitational field'' on its motion through a gravitational plane wave. Indeed, the gravitational self-force is most easily expressed in terms of the tail of an appropriate Green function \cite{GrallaSF, Pound, HarteSF, PoissonRev}. A ``self-torque'' \cite{HarteSF} could be also be computed in this way. More generally, it might be possible to use Penrose limits in order to understand some aspects of the self-force problem on generic backgrounds by considering equivalent problems on plane wave backgrounds. This is especially likely to be relevant for the ``ultra-relativistic'' self-force problem \cite{ChadUltraRel}.

Another potential application for the plane wave Green function $G_{ab}{}^{a'b'}$ involves the long-time behavior of Green functions associated with non-plane wave backgrounds. Within a normal neighborhood, the most singular portion of such a Green function always has the form $U_{ab}{}^{a'b'} \delta_-(\sigma)$. The structure of this term can change over large distances. Penrose limits were used in \cite{HarteCaustics} to derive such changes for scalar Green functions $G$ by reducing the problem to an equivalent one in an appropriate plane wave background. This showed that every encounter with a non-degenerate conjugate point leads to changes in the ``leading-order singularity structure'' of $G$ with the form
\begin{equation}
    |\Delta|^{1/2} \delta(\sigma) \rightarrow \mathrm{pv} \left( \frac{|\Delta|^{1/2}}{ \pi \sigma }  \right) \rightarrow - |\Delta|^{1/2} \delta(\sigma) \rightarrow \ldots
\end{equation}
This pattern is universal. An analogous result for linearized metric perturbations could likely be established if the plane wave Green function of Sec. \ref{Sect:GravGreen} could be extended beyond the normal neighborhood.

\acknowledgments

The author thanks Seth Hopper and Brien Nolan for useful discussions.


\begin{thebibliography}{1}
\bibitem{GuentherBook} P. G\"{u}nther, \textit{Huygen's Principle and Hyperbolic Differential Equations} (Academic Press, San Diego, 1988)
\bibitem{OnlyPlaneWavesEM} P. G\"{u}nther and V. W\"{u}nsch, Math. Nach. \textbf{63}, 97 (1974)
\bibitem{HuygensReview1} M. Belger, R. Schimming, and V. W\"{u}nsch, Z. Anal. Anwend. \textbf{16}, 9 (1997)
\bibitem{HuygensReview2} S. R. Czapor and R. G. McLenaghan, Acta Phys. Polon. B Proc. Supp. \textbf{1}, 77 (2008)
\bibitem{Guenther} P. G\"{u}nther, Arch. Rat. Mech. Anal. \textbf{18}, 103 (1965)
\bibitem{OnlyPlaneWavesScalar} R. G. McLenaghan, Math. Proc. Camb. Phil. Soc. \textbf{65}, 139 (1969)
\bibitem{WeylEqn} V. W\"{u}nsch, Beitr. zur Anal. \textbf{13}, 147 (1979)
\bibitem{Penrose} R. Penrose, p. 271 in \textit{Differential Geometry and Relativity} (Reidel, Dordrecht, 1976)
\bibitem{HarteCaustics} A. I. Harte and T. D. Drivas, Phys. Rev. D \textbf{85}, 124039 (2012)
\bibitem{Hollowood} T. J. Hollowood and G. M. Shore, J. High Energy Phys. \textbf{12}, 091 (2008)
\bibitem{String1} A. A. Tseytlin, Class. Quantum Grav. \textbf{12}, 2365 (1995)
\bibitem{String2} D. Sadri and M. M. Sheikh-Jabbari, Rev. Mod. Phys. \textbf{76}, 853 (2004)
\bibitem{Waylen} P. C. Waylen, Proc. Roy. Soc. Lond. A \textbf{321}, 397 (1971)
\bibitem{GRHuygensNoonan} T. W. Noonan, Astrophys. J. \textbf{343}, 849 (1989)
\bibitem{GRHuygensWuensch} V. W\"{u}nsch, Gen. Relativ. Grav. \textbf{22}, 843 (1990)
\bibitem{NoonanVect} T. W. Noonan, Astrophys. J. \textbf{341}, 786 (1989)
\bibitem{Kuenzle} H. P. K\"{u}nzle, Math. Proc. Camb. Phil. Soc. \textbf{64}, 779 (1968)
\bibitem{EhlersKundt} J. Ehlers and W. Kundt in \textit{Gravitation: An Introduction to Current Research}, edited by L. Witten (Wiley, New York, 1962) 
\bibitem{GlobalLorentz} J. K. Beem, P. E. Ehrlich, and K. L. Easley, \textit{Global Lorentzian Geometry} (Dekker, New York, 1996), 2nd ed.
\bibitem{HarteLensing} A. I. Harte, Class. Quantum Grav. \textbf{30}, 075011 (2013)
\bibitem{Xanth} B. C. Xanthopoulos, J. Math. Phys. \textbf{19}, 1607 (1978)
\bibitem{Bondi} H. Bondi and I. Robinson, Proc. R. Soc. Lond. A \textbf{251}, 519 (1959) 
\bibitem{PlaneWaveClassify} A. Coley, D. McNutt, and R. Milson, Class. Quantum Grav. \textbf{29}, 235023 (2012)
\bibitem{HillBook} W. Magnus and S. Winkler, \textit{Hill's Equation} (Dover, New York, 1979)
\bibitem{PenrosePlane} R. Penrose, Rev. Mod. Phys. \textbf{37}, 215 (1965)
\bibitem{PoissonRev} E. Poisson, A. Pound, and I. Vega, Living Rev. Relativity \textbf{14}, 7 (2011)
\bibitem{CollidingPlane} J. B. Griffiths, \textit{Colliding Plane Waves in General Relativity} (Clarendon Press, Oxford, 1991)
\bibitem{Yurtsever} U. Yurtsever, Phys. Rev. D \textbf{38}, 1731 (1988)
\bibitem{Yurtsever2} U. Yurtsever, Phys. Rev. D \textbf{40}, 329 (1989)
\bibitem{XanthPlane} B. C. Xanthopoulos, J. Math. Phys. \textbf{30}, 2626 (1989)
\bibitem{XanthPlane2} B. C. Xanthopoulos, J. Math. Phys. \textbf{32}, 1866 (1991) 
\bibitem{CosmGreen} P. C. Waylen, Proc. R. Soc. Lond. A \textbf{362}, 233 (1978)
\bibitem{CosmGreen2} R. R. Caldwell, Phys. Rev. D \textbf{48}, 4688 (1993)
\bibitem{Friedlander} F. G. Friedlander, \textit{The Wave Equation on a Curved Spacetime} (Cambridge University Press, Cambridge, 1975)
\bibitem{CurvScalars} A. Coley, S. Hervik, and N. Pelavas, Class. Quantum Grav. \textbf{26}, 025013 (2009)
\bibitem{VSI} V. Pravda, A. Pravdov\'{a}, A. Coley, and R. Milson, Class. Quantum Grav. \textbf{19}, 6213 (2002)
\bibitem{Pound} A. Pound, Phys. Rev. D \textbf{81}, 024023 (2010)
\bibitem{HarteSF} A. I. Harte, Class. Quantum Grav. \textbf{29}, 055012 (2012)
\bibitem{GrallaSF} S. E. Gralla and R. M. Wald, Class. Quantum Grav. \textbf{25}, 205009 (2008)
\bibitem{ChadUltraRel} C. R. Galley and R. A. Porto, arXiv:1302.4486
\end{thebibliography}
\end{document}